\documentclass[letterpaper,twocolumn,10pt]{article}
\usepackage{usenix2,epsfig}
\begin{document}

\date{}

\title{\Large \bf An SDN-based approach to enhance BGP security}

\author{
{\rm  Regivaldo Costa, Fernando M. V. Ramos}\\
LaSIGE, Faculdade de Ci\^encias, Universidade de Lisboa, Portugal\\
rcosta@lasige.di.fc.ul.pt, fvramos@ciencias.ulisboa.pt
} 

\maketitle

\thispagestyle{empty}

BGP, the glue that holds the Internet together, is vulnerable to a series of attacks.
Several incidents demonstrate their recent materialization in the form of prefix hijacks and route leaks~\cite{1}.
The networking community is well aware of the problem, with many solutions having been proposed in the past two decades~\cite{2}.

\textbf{Motivation.} Unfortunately, the most effective solutions for BGP security remain largely undeployed~\cite{1}.
This is fundamentally due to one (or more) of the following reasons.
First, they use cryptographic techniques that are \emph{too computationally expensive} for current routers.
Second, they require \emph{changes to BGP} which, due to its widespread adoption and the lack of a centralized authority to mandate such change, makes migration difficult.
This problem could be overcome if network operators were given \emph{the right incentives} to migrate to a secure solution. 
But this is also hard, particularly as several solutions are not effective until there is wide adoption, while others offer benefits to the Internet as a whole, but not to its adopters. 

\textbf{Contribution.} The separation of the routing control problem from routers as advocated by Software-Defined Networking (SDN) has been proposed as a solution for BGP problems~\cite{3}.
We posit this to be key to address BGP security.
As such, in this abstract we propose an SDN architecture for secure routing.

Our solution --- BGPSecX --- is based on a Software-Defined Internet Exchange (as SDX~\cite{4}, an SDN-based architecture targeting IXPs).
Our design further includes secure channels between the BGPSecX of different IXPs that want to collaborate.
The L2 nature of IXPs make them a good fit for SDN, and its growing importance (they number more than 350 today, and some carry as much traffic as global Tier-1 ISPs~\cite{5}) turns them into a compelling target.
Recently, IXPs have started offering value-added services to their members --- such as Route Server services --- as a successful incentive to join.
We see BGPSecX as an additional value-added service.

BGPSecX will include four security defenses against prefix hijacks.
First, detection by means of ROA verification using a RPKI infrastructure. 
The RPKI provides a trusted mapping from allocated IP prefixes to ASes, and can thus be used to create a whitelist mechanism to filter hijacked routes.  
One problem of the RPKI is coverage, which is still small (less than 5\% of all prefixes are mapped).
This motivates the next techniques.
Second, we include prefix filtering, another whitelist technique used to filter bogus announcements.
The idea is for an AS to keep a list of prefixes announced by its customers (which is usually possible) and discard announcements of prefixes not on the list.  
The main challenge of this technique is the lack of incentives, as this technique does not protect an AS itself, but instead it protects the Internet from attacks by its own customers~\cite{1}.
However, the secure channel will allow IXPs to communicate, and enable the creation of white lists of clusters of IXPs, which can be seen as a more attractive solution.
Third, queries between IXPs will be used for validation of routing information (inspired by~\cite{7}).
Fourth, we will explore the use of well-know anomaly detection mechanisms~\cite{2} and use them at an inter-IXP scale.

We believe our proposal to be capable of addressing the three fundamental problems mentioned in the beginning.
First, by offloading control logic to the SDN controller we remove the computational burden from routers.
Second, our solution does not require changes to the protocol as the techniques only require BGP messages to be intercepted.
Finally, by offering BGPSecX as an additional value-added service from an IXP, we grant incentives for organizations to secure their interdomain routing.  
As IXP members use the service, positive network effects may be created increasing adoption by other members.
Also, techniques that offer little value to the AS that implements them, such as prefix filtering, can gain further relevance at the scale of clusters of IXPs.

\textbf{Implementation.} We implemented a first prototype that mitigates prefix hijacking using a RPKI infrastructure, as a network application for both Floodlight and ONOS.
We are now in the process of validating our initial solution with the use of real BGP traces.

\section*{Acknowledgments}
This project has received funding from the European Union's Horizon 2020 research and innovation programme under grant agreement No H2020-643964 (SUPERCLOUD), and by national funds through Funda\c{c}\~ao para a Ci\^encia e a Tecnologia (FCT) with reference UID/CEC/00408/2013 (LaSIGE).

\end{document}